\documentclass[twocolumn,prb,showpacs]{revtex4}
\usepackage{amsmath}
\usepackage{amssymb}
\usepackage{graphicx}
\usepackage{bm}

\begin{document}

\title{Strain dependent conductivity in biased bilayer graphene}

\author{J. A. Crosse}
\email{alexcrosse@gmail.com}
\affiliation{Department of Electrical and Computer Engineering, National University of Singapore, 4 Engineering Drive 3, Singapore 117583.}

\date{\today}

\begin{abstract}
Intrinsic bilayer graphene is a gapless semimetal. Under the application of a bias field it becomes a semiconductor with a direct band gap that is proportional to the applied field. Under a layer-asymmetric strain (where the upper layer undergoes compression and lower layer tension or visa-versa) we find that the band gap of a biased bilayer graphene ribbon becomes indirect and, for higher strains, becomes negative returning the material its original semimetal state. As a result, the conductivity of the ribbon increases and can be almost an order of magnitude larger that of the intrinsic unbiased material - a change that can be induced with a strain of only $\approx 2-3\,\%$. The conductivity is proportional to the applied strain and the magnitude of the effect is tunable with the bias field. Such layer-asymmetric strains can be achieved by bending, with forces on the order of $\approx 1\,\mathrm{nN}$ resulting in a layer-asymmetric strain of $\approx 1\,\%$. This new electromechanical effect has a wide potential for application in the areas of nano-force microscopy and pressure sensing on the atomic scale.
\end{abstract}


\maketitle

\section{Introduction}

Since its fabrication in 2004 \cite{discov}, single and few layers graphene have been extensively studied and the documentation of this family of materials' remarkable electronic properties \cite{rev1, rev2, rev3} has fuelled speculation as to their role in the next generation of electronic and spintronic devices. Initially, research was predominantly directed towards the single layer variant. However, it's gapless nature is problematic with regard to developing usable graphene transistors \cite{trans} and, although a number of methods have been suggested for gap generation in monolayer graphene \cite{pereiraTBM, confine, dope, substrate}, tailoring the materials electronic properties still remains a major challenge. One alternative solution is to move to the multi-layer variant. The extra layers of bi-, tri- and few layer graphene lead to greater flexibility with regard to band structure engineering. In particular, bilayer graphene (BLG) has been shown to be remarkably versatile. As with single layer graphene it is a gapless semimetal with the Fermi energy running through the Dirac point, though the four bands of BLG are hyperbolic rather than linear \cite{BLG1, BLG2}. However, the bi-layer structure makes it possible to electronically gate the two sides of the material. This bias field changes the relative potential of each layer opening a direct bandgap \cite{bias}, converting BLG from a semimetal to a semiconductor, a more useful electronic configuration for electronic device applications. 

Another method for altering the electronic properties of a material is the use of strain. The band structure of a material is directly related to its crystal lattice. Thus, by applying an external force and distorting the material's crystal structure one can change its electronic response. This approach to band structure engineering has been studied extensively for both mono- and bi- layer graphene. Specifically for the latter, uniform in-plane strains \cite{falko,conductance,peetersblg} and changes in the interlayer distance \cite{peetersblg} have been considered as well as interlayer shearing \cite{maria} and the effect of flexure on the Dirac points \cite{maria}. However, previous work has mainly focussed on unbiased BLG. The effects of a simultaneous bias field and strain have yet to be studied.

Here we report on the prediction that a biased BLG ribbon under a layer-asymmetric strain (where the upper layer undergoes compression and lower layer tension or visa-versa) will first be converted from a direct to an indirect band gap material and, under larger bending strains, will convert back from a semiconductor to a semimetal. The conversion from direct to indirect band gap will lead to longer lifetimes for electron-hole pairs at the band minima since direct recombination no longer conserves momentum and hence is suppressed - a feature that could be useful in applications where carrier lifetime is important. The most notable effect of the semiconductor-semimetal transition is a sharp change in the conductivity with a higher strain leading to a higher conduction. The magnitude of this effect is dependent on the bias field with a stronger bias field leading to a larger change in the conductivity. Such a strain configuration occurs when a material undergoes bending or can be induced by substrate strain by sandwiching the material between substrates with larger and smaller lattice constants, respectively.

\begin{figure}[hbt]
\centering
\includegraphics[width=0.9\linewidth]{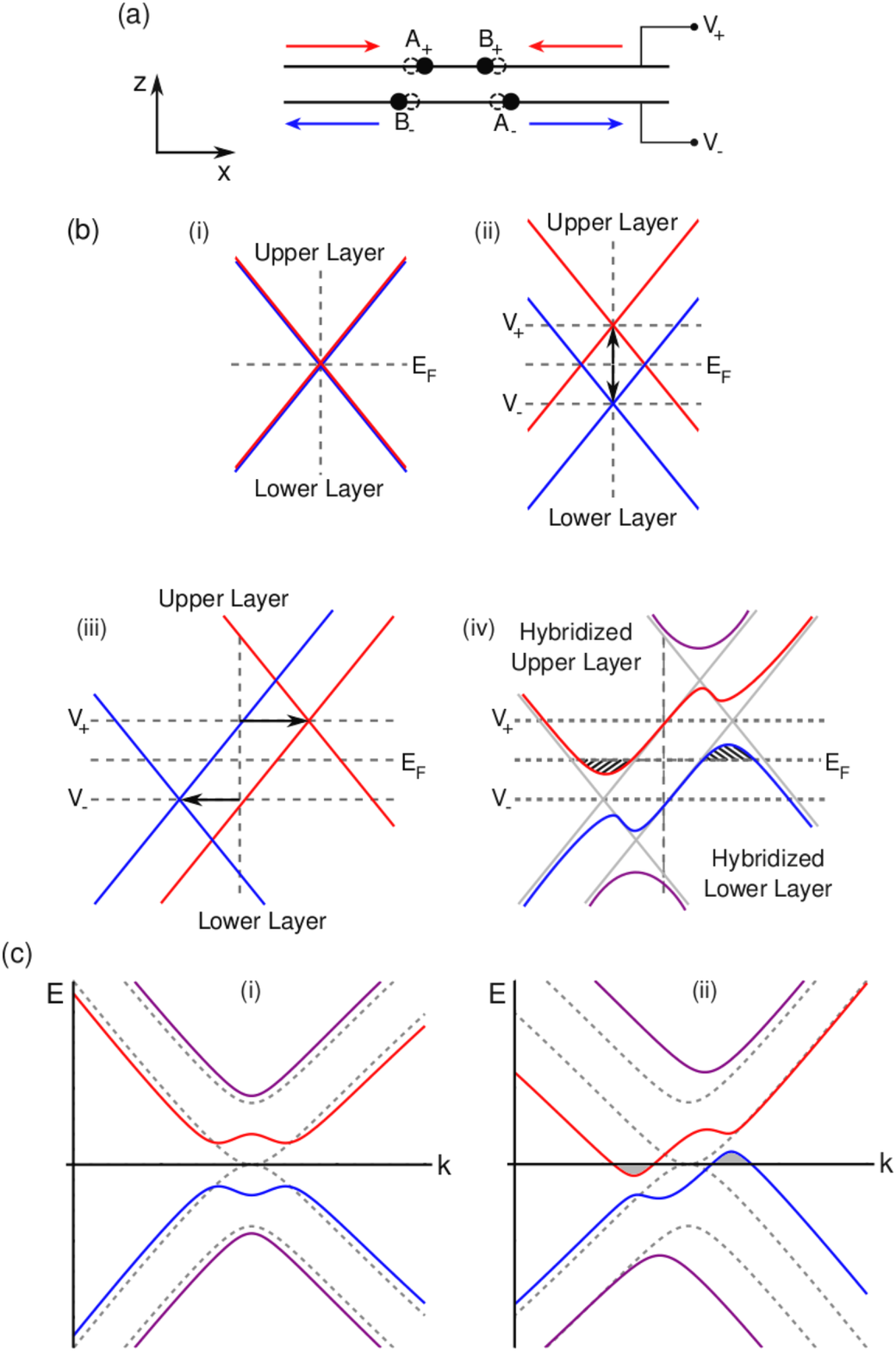}
\caption{(Color online) (a) A layer-asymmetric strained biased graphene bi-layer - the upper layer experiences compression and the lower layer experiences tension. (b) Schematic diagram of the band structure at the Dirac point of biased BLG ribbon under a layer-asymmetric strain. (i) Two uncoupled graphene layers have degenerate Dirac cones. (ii) Application of equal an opposite potentials $\pm V$ to the two layers breaks the degeneracy moving one to a lower energy and one to a higher energy. (iii) Application of compressive and tensile strain to opposing layers moves one Dirac cone to higher momenta and the other to lower momenta. (iv) The van der Waals coupling that bonds the two layers of the BLG ribbon hybridizes the two Dirac cones leading to a `skewed double well' topology for the bands which if severe will convert the the ribbon from a semiconductor to a semimetal. (c) The band structure of a unbiased (dashed) and biased BLG (solid) in the absence of a layer-asymmetric strain (i) and in the presence of a layer-asymmetric strain (ii).}
\label{diag}
\end{figure}

A brief qualitative analysis illustrates the physical mechanism behind the effect. Consider a BLG ribbon subject to compression on the upper layer and tension on the lower layer [see Fig. \ref{diag} (a)]. Compression narrows the unit cell of the crystal lattice and hence stretches the Brillouin zone. As a result, the Dirac points are move outwards to higher momenta. Tension, on the other hand, lengthens the units cell and hence the brillouin zone contracts moving the Dirac points to lower momenta [see Fig. \ref{diag} (b)]. This opposing shift of the Dirac points of the two layers leads to `$k$-space skewing' of the bands, generating the indirect gap. At high strains, the conduction band is pushed below and the valence band rises above the Fermi energy, generating a carrier population in the bands and a non-vanishing conductivity [see Fig. \ref{diag} (c)]. 

The rest of the paper will be devoted to a quantitative analysis of this effect. Firstly, we relate the applied strain to the changes in the electronic properties of the material by computing the electronic band structure of a strained BLG ribbon. To do this, we substituting the strain tensor for the deformation into the microscopic tight-binding model for BLG, a well known technique from previous studies of strained graphene \cite{pereiraTBM,pereiraA,error,peeters2,domain}. From the band structure it is possible to derive the electronic properties of the material. Secondly, we will discuss the ability to generate such strains via bending.

\section{Band Structure of BLG subject to a Layer Asymmetric Strain}

\begin{figure}[b]
\centering
\includegraphics[width=0.9\linewidth]{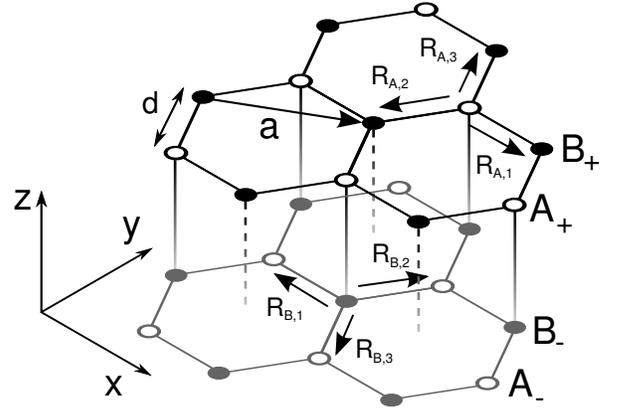}
\caption{The bilayer graphene lattice.}
\label{lat}
\end{figure}

The crystal structure of BLG consists two monolayer graphene sheets weakly bound via inter-layer Van der Waals forces. The two sheets exhibit Bernal stacking where the $A$ sublattice of the upper (+) sheet aligned vertically with the $B$ sublattice of the lower (-) sheet [see Fig. \ref{lat}]. The length of the in-plane lattice vector is $a = 2.46\mathrm{\AA}$ and the nearest-neighbour vectors read
\begin{subequations}
\begin{align}
\mathbf{R}_{A,1} &= \left(\begin{array}{c}
\frac{a}{\sqrt{3}}\\
0
\end{array}\right), \hspace{0.2cm} 
\mathbf{R}_{A,2} = \left(\begin{array}{c}
-\frac{a}{2\sqrt{3}}\\
\frac{a}{2}
\end{array}\right), \hspace{0.2cm} 
\mathbf{R}_{A,3} = \left(\begin{array}{c}
-\frac{a}{2\sqrt{3}}\\
-\frac{a}{2}
\end{array}\right),\\
\mathbf{R}_{B,1}& = \left(\begin{array}{c}
-\frac{a}{\sqrt{3}}\\
0
\end{array}\right), \hspace{0.2cm} 
\mathbf{R}_{B,2} = \left(\begin{array}{c}
\frac{a}{2\sqrt{3}}\\
\frac{a}{2}
\end{array}\right), \hspace{0.2cm} 
\mathbf{R}_{B,3} = \left(\begin{array}{c}
\frac{a}{2\sqrt{3}}\\
-\frac{a}{2}
\end{array}\right),
\end{align}
\end{subequations}
each with length $d=a/\sqrt{3} = 1.42\,\mathrm{\AA}$.

In order to evaluate the band structure of BLG under a layer-asymmetric strain we use a nearest-neighbour tight binding model \cite{peetersblg}. Previous studies have considered higher order inter-layer hopping terms in the Hamiltonian \cite{maria, blgrev}. This leads to added structure near the Dirac point - the single parabolic minimum splits to give four Dirac cones. However, the energy scale of these features is small, $\approx 1\,\mathrm{meV}$, and hence are only observable in the most pristine of BLG samples where impurity generated disorder has not obscured them \cite{maria}. Thus, for most applications, where the BLG samples are not sufficiently free of impurities or the energy scale of the effect is greater than a few $meV$, the nearest-neighbour hopping suffices. Furthermore, one finds that the application of a bias field with an energy $>1\,\mathrm{meV}$, as will be the case here, will also obscures these features. Thus, in the following, we will consider only the nearest neighbour inter-layer hopping term. A brief discussion of the effect of the higher order inter-layer hopping terms can be found in Appendix \ref{app:hot}.  

We consider a strained BLG ribbon subject to bias field of $\pm V$ applied to the upper and lower layers, respectively. The ribbon is orientated such that the armchair edge is parallel to the $x$-axis (results for other orientations can be found by a simple rotational transformation of the strain tensor) and is of sufficient size that confinement effects are negligible (i.e. the band structure can be considered to be that of bulk graphene). The nearest-neighbour tight-binding Hamiltonian for this system reads \cite{BLG1, BLG2}
\begin{equation}
\hat{H} = \left(\begin{array}{cccc}
V & f^{+}_{A}(\mathbf{k}) & 0 & 0 \\
f^{+}_{B}(\mathbf{k}) & V & 2\gamma & 0 \\
0 & 2\gamma & -V & f^{-}_{A}(\mathbf{k}) \\
0 & 0 & f^{-}_{B}(\mathbf{k}) & -V \\
\end{array}\right),
\label{H}
\end{equation}
with $\gamma$ the interlayer coupling and the electron hopping phase factor reading
\begin{equation}
f^{\pm}_{i}(\mathbf{k}) = \sum_{\alpha}t_{\alpha}e^{-i\mathbf{k}\cdot[\mathbf{R}_{i,\alpha} + \bm{\Omega}_{\pm}\cdot\mathbf{R}_{i,\alpha}]},
\label{f}
\end{equation}
where the sum over $\alpha$ is the sum over all the in-plane nearest-neighbour vectors. Here, $\bm{\Omega}_{\pm}$ is the strain-induced transformation of the in-plane nearest-neighbour vectors. The ribbon is subject to uniaxial compression in the $x$ direction in upper layer and equal (but opposite) uniaxial tension in the $x$ direction in the lower layer, hence \cite{pereiraTBM}
\begin{equation}
\bm{\Omega}_{\pm} = \pm \varepsilon
\left(\begin{array}{cc}
-1 & 0\\
0 & \nu
\end{array}\right),
\end{equation}
where $\nu$ is the poisson ratio of graphene. (The derivation of $\bm{\Omega}_{\pm}$ for a general bending strain can be found in Appendix \ref{app:ebt} and for the specific case of a layer-asymmetric strain induced by a central point load in Appendix \ref{app:scl}.) The change in bond length also results in a change in the hopping potentials. Here, $t_{\alpha} = t_{0}\,\mathrm{exp}\left[-\beta\left(l/d-1\right)\right]$ is the renormalized hopping amplitude \cite{pereiraTBM} with $d$ the unstrained bond length, $l = |\mathbf{R}_{\alpha} + \bm{\Omega}_{\pm}\cdot\mathbf{R}_{\alpha}|$ the strained bond length, $t_{0}$ the unstrained hopping amplitude and $\beta$ the hopping decay parameter. In principle, the opposing strains of in upper and lower layer will increase the distance between the $A_{+}$ and $B_{-}$ atoms thus changing the interlayer coupling $\gamma$. However, for even for strains of $\approx 10\,\%$, the maximum discussed here, the change in this distance is $\approx 0.06\,\%$ and hence is negligible.

The standard diagonalization of Eq. \eqref{H} and the usual expansion about the Dirac points at $\mathbf{K}_{\pm}=(0,\pm 4\pi/3\sqrt{3}d)$ leads to the low energy band structure of the bilayer graphene ribbon as a function of the applied bias field, $V$, and the strain, $\varepsilon$. We evaluate the band structure for a graphene ribbon subject to a strain of up to $6\%$ and a bias field of up to $0.3\,\mathrm{eV}$. The unstrained hopping amplitude\cite{saito} is taken to be $t_{0}=3$, the poisson ratio \cite{poisson} $\nu=0.165$, the interlayer coupling \cite{gamma} $\gamma = 0.202\,\mathrm{eV}$ and the hopping decay parameter \cite{pereiraTBM} $\beta=3$.

\section{Band Gap}

\begin{figure}[htb]
\centering
\includegraphics[width=0.9\linewidth]{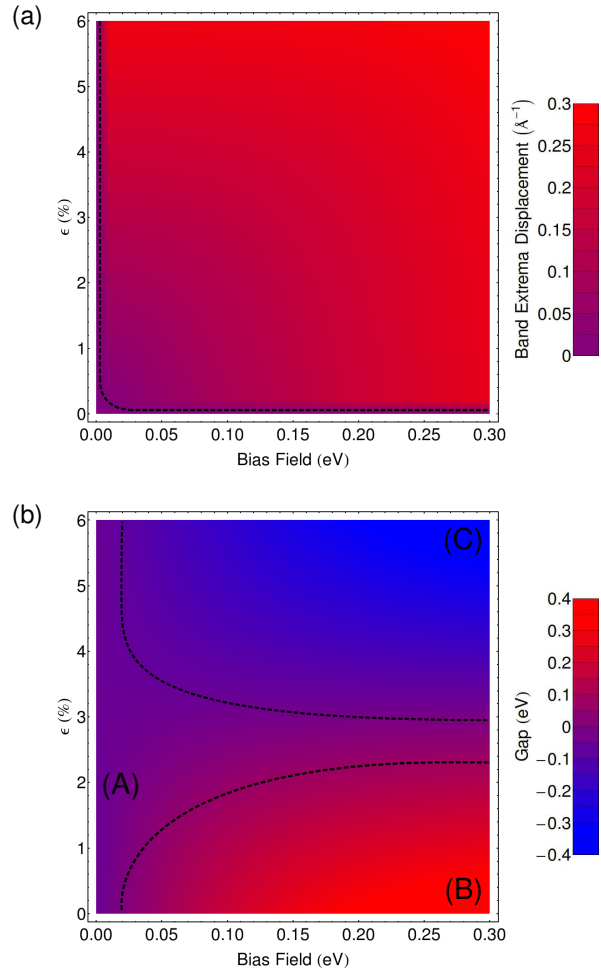}
\caption{(Color online) (a) The difference in momenta of the conduction and valence band extrema as a function of bias voltage and applied strain. Below and to the left of the dashed line difference is zero hence the gap is direct. Above and to the right of dashed line the difference in momenta is non-zero hence the gap is indirect. (b) The band gap as a function of bias voltage and applied strain. In the region marked (A) the BLG ribbon is a low conductivity semimetal - gap is small or vanishing and the intrinsic carrier population is low. In the region marked (B) the BLG ribbon is a semiconductor - the gap is large and the intrinsic carrier population vanishes. In the region marked (C) the BLG ribbon is a high conductivity semimetal - the gap is negative and the intrinsic carrier population is high.}
\label{indgap}
\end{figure}
In the absence of both a bias field and strain, bilayer graphene is a gapless semimetal. On the application of a bias field the potential difference between the upper and lower layer opens a gap and the material changes from a semimetal to a semiconductor. The band structure takes a `double well' form with two degenerate minima on either side of the Dirac point [see Fig. \ref{diag}c (i)]. Application of a layer-asymmetric strain results in compression in the upper layer, shifting the band structure to larger $|\mathbf{k}|$, and tension in the lower layer, shifting the band structure to smaller $|\mathbf{k}|$. This causes one of the conduction band minima to rise in energy and the other to fall. Similarly, the strain causes the corresponding valence band maxima to rise in energy and the other to fall. This breaks the degeneracy of the band extrema and results in the sudden generation of an indirect band gap [see Fig. \ref{diag}c (ii)]. The difference in $\mathbf{k}$ of the conduction and valence band extrema increases with both strain and bias field. This effect is shown in Fig. \ref{indgap} (a). 

The change from a direct to indirect band gap will have a significant effect on the electronic properties of the ribbon. Most notably the recombination lifetime of electron and hole pairs at the band minima will be greatly enhanced since phonon exchange with the crystal lattice is required to conserve momentum. This will be a great advantage for applications which require long carrier lifetimes.

\section{Conductivity}

\begin{figure}[b]
\centering
\includegraphics[width=0.9\linewidth]{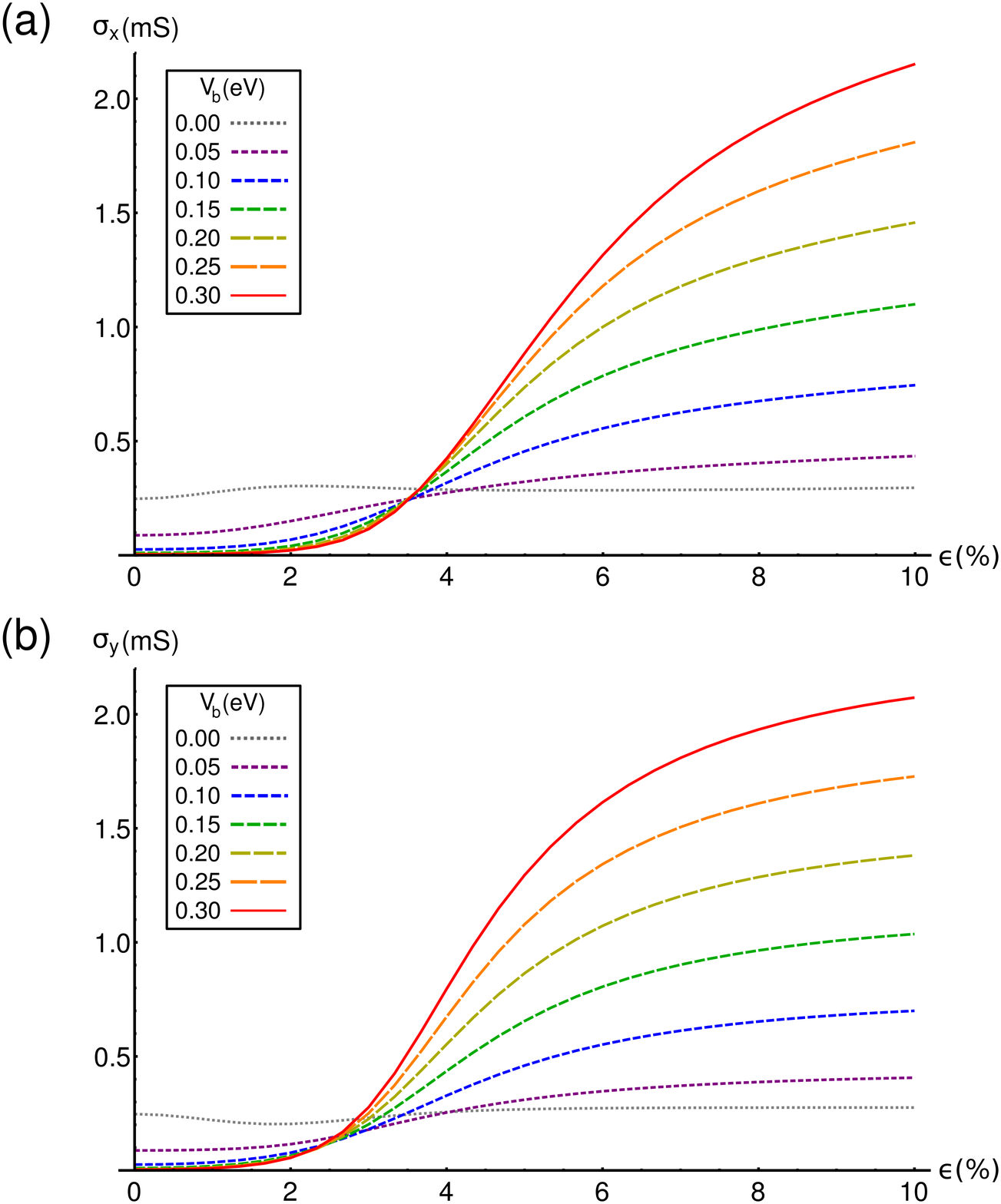}
\caption{(Color online) The (a) $x$- and (b) $y$- conductivity as a function of applied strain for different bias fields.}
\label{cond}
\end{figure}

If the applied layer-asymmetric strain is increased further the band extrema will continue to rise/fall, narrowing the gap. At high enough strains the conduction band falls below, and the valence band rises above, the Fermi energy. At this point the gap is negative and one generates a population of carriers in both the conduction and valence bands. Hence, the ribbon changes from a semiconductor back to a semimetal. This effect is shown in Fig. \ref{indgap} (b). 

The change from semiconductor to semimetal can be quantified by computing the conductivity of the BLG ribbon as a function of strain. The conductivity tensor, $\bm{\sigma}$, is defined in terms of the current density $\mathbf{j}$ and applied electric field $\mathbf{E}$ by $\mathbf{j} = \bm{\sigma}\cdot\mathbf{E}$ and can be found by considering the change in the equilibrium carrier distribution, $f(E)$, under a perturbation, a change which, in turn, can be found from the Boltzmann equation
\begin{equation}
\left.\frac{\partial f(E)}{\partial t} + \mathbf{v}_{f}(\mathbf{k})\cdot\frac{\partial f(E)}{\partial \mathbf{r}} +\frac{\partial \mathbf{k}}{\partial t}\cdot\frac{\partial f(E)}{\partial \mathbf{k}} = \frac{\partial f(E)}{\partial t}\right|_{scat}.
\label{boltzmann}
\end{equation}
Here, $\mathbf{v}_{f}(\mathbf{k}) = \nabla_{\mathbf{k}}E(\mathbf{k})/\hbar$ denotes the Fermi velocity and the equilibrium carrier distribution is given by the Fermi-Dirac distribution
\begin{equation}
f(E) = \frac{1}{1+\mathrm{exp}\left[\frac{E-E_{F}}{k_{b}T}\right]}.
\end{equation}
In the relaxation time approximation, which assumes that the carrier distribution returns to equilibrium in a uniform manner, one can rewrite the scattering term on the R.H.S of Eq. \eqref{boltzmann} as $-\delta f(E)/\tau$, where $\delta f(E)$ is the difference of the carrier distribution from the equilibrium distribution and $\tau$ is the carrier relaxation time, which characterizes how fast the perturbed carrier distribution returns to equilibrium. Furthermore, for d.c. fields in the absence of thermal gradients the carrier distribution displays no explicit temporal or spatial dependence and hence the first two terms on the L.H.S vanish. For small electric fields the remaining term in Eq. \eqref{boltzmann} can linearize, resulting in a simple expression for the $2D$ conductivity tensor
\begin{equation}
\bm{\sigma} \approx -\frac{e^2\tau}{\pi^{2}}\sum_{v,c}\int d^{2}k\,\mathbf{v}_{f}(\mathbf{k})\otimes\mathbf{v}_{f}(\mathbf{k})\frac{\partial f(E)}{\partial E}.
\end{equation}
The summation indicates that contributions from both the valence and conduction band need to be included and the extra factor of $4$ comes from the spin and valley degrees of freedom.

Figure \ref{cond} shows the $x$ and $y$ components of the conductivity tensor as a function of strain for various bias fields (note that the off diagonal components, $\sigma_{xy}$, vanish). The relaxation rate was taken to be $\tau = 1.3\,ps$, which is comparable to values obtained in current relaxation time measurements \cite{sun, george} and results in an intrinsic conductivity of unbiased BLG similar to experimentally measured values \cite{bias}. The temperature was taken to be $T=300\,K$. In the absence of a bias field the bilayer graphene ribbon is a semimetal and, hence, at finite temperature, has a minimum conductivity which is independent of the strain. As the bias field increases a gap opens and, for small strains, one sees a reduction in the conductivity with respect to the zero field case. However, as the applied strain is increased, the gap narrows and the conductivity grows. For high strains the gap is negative and there is an intrinsic carrier population in the conduction and valence bands. Hence, the conductivity becomes significantly higher than the zero bias field configuration. The slight variation in $\sigma_{x}$ compared to $\sigma_{y}$ is owing to slightly different Fermi velocities in the $x$- and $y$-directions near the Dirac points. This variation is a result of the strain which, as it is uniaxial, has a minor directional effect. We note that a similar effect has been investigated in monolayer graphene \cite{physchem,low,vanveen}, however the tunable sensitivity of the effect in bilayer graphene makes it a more versatile for applications.

\section{Layer Asymmetric Strain via Bending}

One way to achieve a layer asymmetric strain is via bending. Under a downward acting bending force the upper surface will undergo compression and the lower surface will experience tension. One method for quantitatively analysing bending is Euler-Bernoulli beam theory, which is a form of linear elasticity theory that aims to describe the deflection of beams subject to a lateral load (see Appendix \ref{app:ebt}). The main assumption is that shear and rotational effects are small and hence is applicable to beams where the length is much greater than the thickness. Thus, it is an appropriate approach to BLG where the typical ribbon length is much greater than the thickness of the two atomic layer.

Consider a simply-supported graphene ribbon of length $L_{x}$, width $L_{y}$ and thickness $L_{z}$. The origin is taken to be at the centre of the beam on the neutral axis and, hence, the upper and lower graphene layers are located at $z =\pm L_{z}/2$. A line load of total force $F_{0}$ acting in the negative $z$ direction is applied to the centre of the beam along the line $x=0$. A straight-forward calculation (see Appendix \ref{app:scl}) yields an in-plane strain of
\begin{equation}
\varepsilon = \frac{3F_{0}L_{x}}{2EL_{z}^{2}L_{y}},
\end{equation}
where $E$ is the Young's modulus of graphene. Furthermore, the force induces a deflection that is maximal at $x=0$, the midpoint of the beam. The magnitude of this deflection is given by
\begin{equation}
w_{max} = \frac{F_{0}L_{x}^{3}}{4EL_{y}L_{z}^{3}} = \frac{L_{x}^{2}}{6L_{z}}\varepsilon.
\label{deflect}
\end{equation}

In the interest of practicality one would ideally like to optimise the strain whilst reducing the deflection. From Eq. \eqref{deflect} one sees that short, thick beams are preferable. For a BLG ribbon with $L_{x} = 100\,\mathrm{nm}$ and $L_{y} = 25\,\mathrm{nm}$ (which is experimentally achievable \cite{mohanty}) and the thickness taken to be $L_{z} = 7\,\mathrm{\AA}$, twice the thickness of monolayer graphene \cite{youngs}, and a Young's modulus of $E =1\,\mathrm{TPa}$ \cite{youngs}, one finds that a $1\,\mathrm{nN}$ force results in a strain of $\approx 1.2\,\%$. This is comparable to the experimentally measured strains found when monolayer graphene undergoes bending \cite{pet}. This strain profile is accompanied by a deflection of $\approx 30\,nm$ ($\approx 30\,\%$). Although this seems like a large deflection, deflections of $\approx 12\%$ have been studied in bulk monolayer graphene \cite{peeters} and deflections of $\approx 20\%$ have been studied in BLG \cite{nanoindent}. The reason for both the large deflection and the low strains is small thickness, $L_{z}$, of the BLG ribbon. From Eq. \eqref{deflect} one sees that the maximum deflection (for fixed strain) scales as $w_{max} \propto 1/L_{z}$. Thus using few layer graphene rather than BLG would reduce the deflection (gap opening by a bias field has already been observed in both tri-layer and few-layer graphene with certain crystalline structures \cite{trilayer, fewlayer}). However, the fundamental physical principle behind the effect is identical to the bilayer treatment presented here.

\section{Summary}

We have shown that a biased BLG ribbon generates an indirect band gap and changes from a semiconductor to a semimetal under a layer-asymmetric strain. The conductivity of the BLG ribbon is proportional to the strain and the magnitude of the effect tunable by the bias field. Such strain profiles can be induced by bending with forces on the order of $1\,\mathrm{nN}$. Thus, this mechano-electronic effect has great potential for applications where nanoscale forces need to be measured; the sensitivity of the BLG ribbon lends itself atomic scale pressure sensors. It is envisaged that this effect will a play key role in development of nano-electromechanical graphene based devices.

\section{Acknowledgements}

The author would like to thank P. Del Linz for useful discussions and A. Danner and the members of the Optical Device Research Group at the National University of Singapore for their hospitality.

\appendix

\section{Higher Order Inter-Layer Hopping Terms}
\label{app:hot}

\begin{figure}[htb]
\centering
\includegraphics[width=0.9\linewidth]{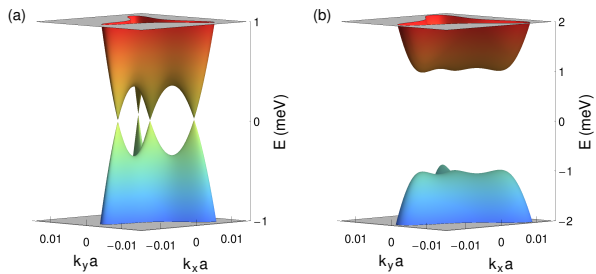}
\caption{(Color online) The band structure near the Dirac point when the higher-order inter-layer hopping terms are taken into account for $\varepsilon = 0\,\%$ and (a) $V=0\,\mathrm{meV}$ and (b) $V=1\,\mathrm{meV}$.}
\label{HOBV}
\end{figure}

The band structure of BLG near the Dirac point is parabolic when only the nearest neighbour inter-layer hopping term, $\gamma$, ($A_{+}-B_{-}$) is considered. This provides and accurate description for electrons with energies $> 1-2\,\mathrm{meV}$ or electrons not too close to the Dirac point. If one includes the higher order inter-layer hopping terms, $\gamma_{3}$ ($B_{+}-A_{-}$) and $\gamma_{4}$ ($A_{+}-A_{-}$ and $B_{+}-B_{-}$), the parabolic Dirac point splits into four Dirac cones \cite{blgrev,maria}. This added structure results in novel physics for low energy electrons close to the Dirac point and although this regime requires low temperatures and very high quality material samples it is, nonetheless, experimentally accessible \cite{mayo}. Including these higher-order terms in the Hamiltonian leads to
\begin{equation}
\hat{H} = \left(\begin{array}{cccc}
V & f^{+}_{A}(\mathbf{k}) & \gamma_{4}f^{+}_{A}(\mathbf{k}) & \gamma_{3}f^{+}_{B}(\mathbf{k}) \\
f^{+}_{B}(\mathbf{k}) & V & 2\gamma & \gamma_{4}f^{+}_{A}(\mathbf{k}) \\
\gamma_{4}f^{+}_{B}(\mathbf{k}) & 2\gamma & -V & f^{-}_{A}(\mathbf{k}) \\
\gamma_{3}f^{+}_{A}(\mathbf{k}) & \gamma_{4}f^{+}_{B}(\mathbf{k}) & f^{-}_{B}(\mathbf{k}) & -V \\
\end{array}\right),
\label{HHOB}
\end{equation}
with the usual electron hopping phase factors, $f^{\pm}_{i}(\mathbf{k})$ given in Eq. \eqref{f}. In the following we take the higher-order hopping parameters to be $\gamma_{3}=178\,\mathrm{meV}$ and $\gamma_{4}=25\,\mathrm{meV}$, which were calculated from the Fermi velocities given in the literature \cite{maria}. As with $\gamma$, the change in the interlayer distance is an order of magnitude smaller than in-plane strains and hence we assume that $\gamma_{3}$ and $\gamma_{4}$ are independent of the strain. In the absence of a bias field and strain, diagonalization leads to the expected four Dirac cones with structure on the energy scale of $\approx 1\,\mathrm{meV}$ [See Fig. \ref{HOBV} (a)]. On application of a bias field a gap opens and one sees that for fields strengths larger than the energy scale of the four Dirac cones the detailed structure of the Dirac point is suppressed [See Fig. \ref{HOBV} (b)]. Since, in the main text, we consider bias fields two orders of magnitude higher (on the order of $\approx 100\,\mathrm{meV}$) the higher-order inter-layer hopping terms do not affect the main results.

\begin{figure}[htb]
\centering
\includegraphics[width=0.9\linewidth]{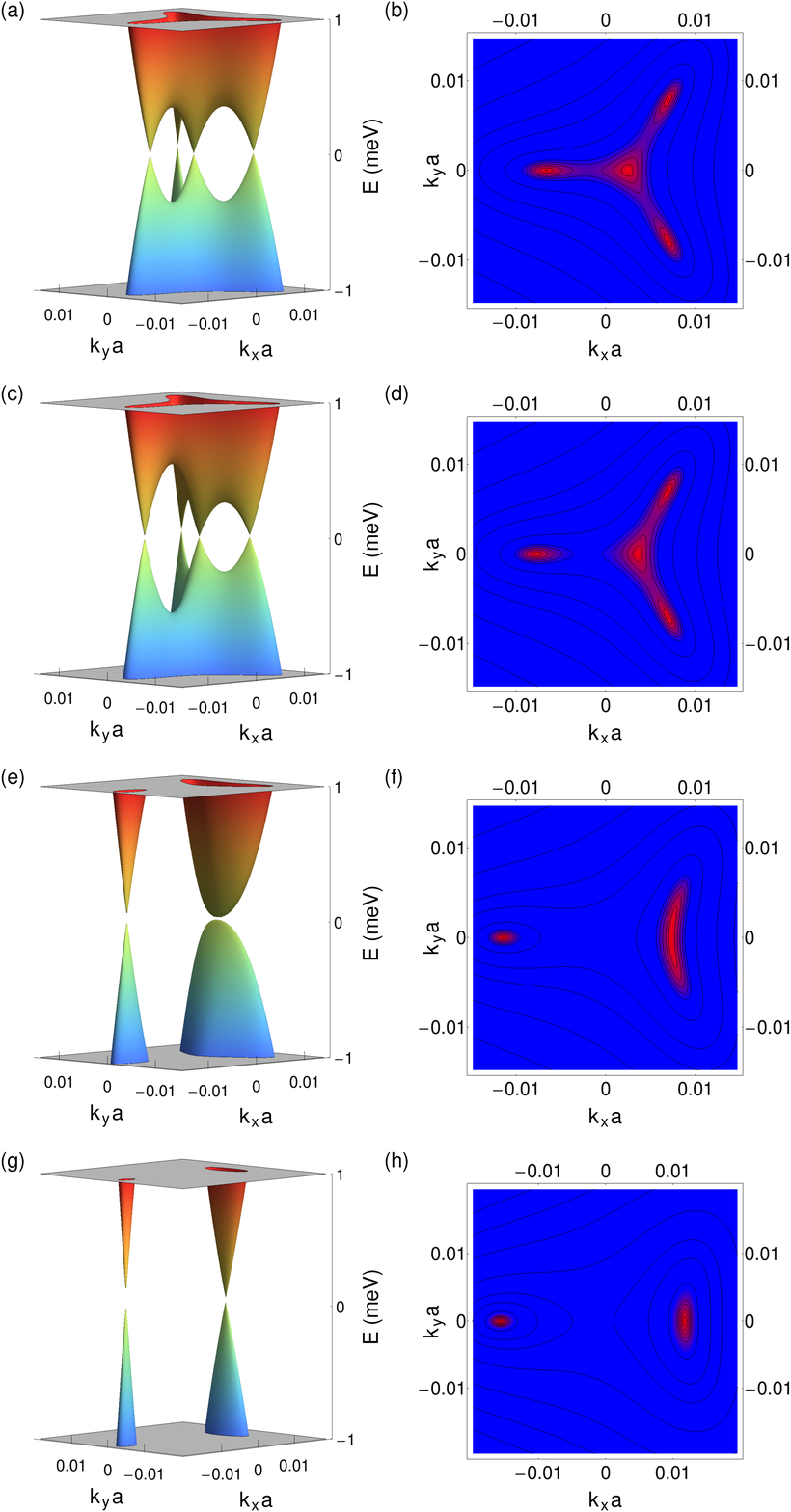}
\caption{(Color online) The band structure and valence band contour plot near the Dirac point when the higher-order inter-layer hopping terms are taken into account for $V=0\,\mathrm{meV}$ and (a) and (b) $\varepsilon = 0$, (c) and (d) $\varepsilon = 0.1\,\%$, (e) and (f) $\varepsilon = 0.3\,\%$ and (g) and (h) $\varepsilon = 0.5\,\%$.}
\label{HOBE}
\end{figure}

For the situation where there is no bias field the low energy features of the Dirac point can be observed [See Fig. \ref{HOBE} (a) and (b)].  Application of a layer-asymmetric strain on the order of $\varepsilon \approx 0.1\,\%$ alters the band structure [See Fig. \ref{HOBE} (c) and (d)]. As one increase the strain the central Dirac point merges with two others. The fourth Dirac point is pushed to higher momenta [See Fig. \ref{HOBE} (e) and (f)]. For strains of $\varepsilon \approx 0.5\,\%$ the merged Dirac point narrows and a gap opens at the fourth Dirac point [See Fig. \ref{HOBE} (g) and (h)].

\section{Euler-Bernoulli Beam Theory}
\label{app:ebt}

For a beam which is extended in the $x$ direction, has width in the $y$ direction and thickness in the $z$ direction, the deflection can be related to the applied distributed force, $F(x)$, which acts in the $z$ direction, via
\begin{equation}
\frac{d^{2}}{dx^{2}}\left(EI\frac{d^{2}w(x)}{dx^{2}}\right) = F(x),
\end{equation}
where $w(x)$ is the deflection of the beam, $E$ is the Young's modulus and $I$ is the second moment of area, a geometric factor that accounts for the $y-z$ cross-sectional shape of the beam. From the deflection the various stresses in the beam can be calculated. The in-plane tensile stress, $s_{xx}(x)$, is given by
\begin{equation}
s_{xx}(x) = -zE\frac{d^{2}w(x)}{dx^{2}} = \frac{Mz}{I},
\label{sxx}
\end{equation}
where $z$ is the distance from the neutral axis (the plane of vanishing stress) and $M$ is the bending moment, which is related to the strength and relative location of the applied force. All other in-plane stress components are assumed to vanish.

Given the components of the stress tensor the components of the strain tensor can be found from the generalized Hookes law, $\varepsilon_{ij} = S_{ijkl}\sigma_{kl}$. The rank-4 tensor $S_{ijkl}$ is the compliance tensor whose components are related to the mechanical properties of the material. 
For the in-plane components, BLG acts like a planar, isotropic material, and hence the transformation reads
\begin{equation}
\left(\begin{array}{c}
\varepsilon_{xx}(x)\\
\varepsilon_{yy}(x)\\
\varepsilon_{xy}(x)\\
\end{array}\right) = 
\frac{1}{E}\left(\begin{array}{ccc}
1 & -\nu & 0 \\
-\nu & 1 & 0\\
0 & 0 & (1+\nu)
\end{array}\right)
\left(\begin{array}{c}
s_{xx}(x)\\
s_{yy}(x)\\
s_{xy}(x)\\
\end{array}\right),
\label{trans}
\end{equation}
where $\nu$ is the poisson ratio.

The strain tensor components in Eq. \eqref{trans} are those for the symmetric infinitesimal strain tensor and hence describe small, spatially constant, rotation free strains. One can find the rotational contribution to the strain by computing the antisymmetric rotational tensor \cite{error,peeters2,domain}, $\bm{\omega}$, which is related to the symmetric infinitesimal strain tensor, $\bm{\varepsilon}$, via $\bm{\nabla}\times\bm{\omega}=-\bm{\nabla}\times\bm{\varepsilon}$. 

In addition, strain contributions owing to the spatial variation of the strain tensor can be included by computing the finite displacement term \cite{domain}. This term is obtained by integrating the metric connections, $\Gamma_{ijk}$, of the infinitesimal strain tensor over the deformation. Thus, the finite displacement term, $\bm{\Sigma}(\mathbf{R})$, at $\mathbf{R}$ reads
\begin{equation}
\bm{\Sigma}(\mathbf{R}) = \int_{0}^{1} d\lambda\,\lambda\Gamma_{ijk}[\mathbf{R}(\lambda)]R_{j},
\end{equation}
where $\mathbf{R}(\lambda) = \lambda \mathbf{R} = (\lambda R_{x},\lambda R_{y},\lambda R_{z})$ and
\begin{equation}
\Gamma_{ijk} = \frac{1}{2}\left(\frac{\partial g_{ij}}{\partial x_{k}}+\frac{\partial g_{ik}}{\partial x_{j}}-\frac{\partial g_{jk}}{\partial x_{i}}\right),
\label{con}
\end{equation}
with the infinitesimal unit of length given by $dl^{2} = g_{ij}dR_{i}dR_{j} = (\delta_{ij}+2\varepsilon_{ij})dR_{i}dR_{j}$. This term accounts for large strains that go beyond the usual infinitesimal strain theory. 

Finally, from the above strain tensors one can find the change in the interatomic distance, $\mathbf{R}_{\alpha}$, of the atoms in a crystal lattice \cite{domain} 
\begin{equation}
\mathbf{R}_{\alpha} \rightarrow \mathbf{R}_{\alpha} + \bm{\Omega}(\mathbf{R})\cdot\mathbf{R}_{\alpha}
\end{equation}
where $\mathbf{R}$ is the location of the atom and
\begin{equation}
\bm{\Omega}(\mathbf{R}) = \bm{\varepsilon}(\mathbf{R}) + \bm{\omega}(\mathbf{R}) - 2\bm{\Sigma}(\mathbf{R}).
\label{Omega}
\end{equation}
Thus, using the above method, it is possible to relate an applied bending force to the bond deformation within the material.

\section{The Strain Tensor For A Central Load}
\label{app:scl}

In the following, we consider a graphene ribbon of length $L_{x}$, width $L_{y}$ and thickness $L_{z}$. The origin is taken to be at the centre of the beam on the neutral axis and, hence, the upper and lower graphene layers are located at $z =\pm L_{z}/2$. A line load of total force $F_{0}$ acting in the negative $z$ direction is applied to the centre of the beam along the line $x=0$. Thus, $F(x)=F_{0}\delta(x)$. From this expression for the force one finds that the bending moment has the form $M=F(|x|-L_{x}/2)/2$. The second moment of area of a rectangular beam reads $I=L_{y}L_{z}^{3}/12$. Substituting these expressions into Eq. \eqref{sxx} and applying the transformation in Eq. \eqref{trans} gives
\begin{equation}
\bm{\varepsilon}_{\pm} = \pm \frac{3F_{0}}{2EL_{z}^{2}L_{y}}
\left(\begin{array}{cc}
\left(2|x|-L_{x}\right) & 0\\
0 & -\nu\left(2|x|-L_{x}\right)
\end{array}\right),
\label{strain}
\end{equation}
for the infinitesimal strain tensor. As the $x$ and $y$ tensile stresses vary with position one also finds a non-zero rotation \cite{error,peeters2,domain}
\begin{equation}
\bm{\omega}_{\pm} = \pm \frac{3F_{0}}{2EL_{z}^{2}L_{y}}
\left(\begin{array}{cc}
0 & -2\nu\,\mathrm{sgn}[x]y\\
2\nu\,\mathrm{sgn}[x]y & 0
\end{array}\right),
\end{equation}
and a finite displacement term \cite{domain}
\begin{equation}
\bm{\Sigma}_{\pm} = \pm \frac{3F_{0}}{2EL_{z}^{2}L_{y}}
\left(\begin{array}{cc}
|x| & \nu\,\mathrm{sgn}[x]y\\
-\nu\,\mathrm{sgn}[x]y & -\nu |x|
\end{array}\right).
\end{equation}
Thus, from Eq. \eqref{Omega}, one finds that the change in the interatomic distance is given by
\begin{equation}
\bm{\Omega}_{\pm} = \pm \frac{3F_{0}L_{x}}{2EL_{z}^{2}L_{y}}
\left(\begin{array}{cc}
-1 & -4\nu\,\mathrm{sgn}[x]y/L_{x}\\
4\nu\,\mathrm{sgn}[x]y/L_{x} & \nu
\end{array}\right).
\end{equation}
Note that the dominant effect of bending is strain in the $x$ direction. As the poisson ratio of graphene \cite{poisson} is $0.165$, all other contributions are a factor of $\approx 6$ less. For ribbons with large aspect ratios the off diagonal terms become negligible.

One final consideration is the maximum deflection, $w_{max}$, of the graphene ribbon under these conditions. This is found to be
\begin{equation}
w_{max} = \frac{F_{0}L_{x}^{3}}{4EL_{y}L_{z}^{3}}.
\end{equation}



\begin{thebibliography}{99}

\bibitem{discov}
K. S. Novoselov, A. K. Geim, S. V. Morozov, D. Jiang, Y. Zhang, S. V. Dubonos, I. V. Grigorieva and A. A. Firsov, Science \textbf{306}, 666 (2004).

\bibitem{rev1}
A. H. Castro Neto, F. Guinea, N. M. R. Peres, K. S. Novoselov and A. K. Geim, Rev. Mod. Phys. \textbf{81}, 109 (2009).

\bibitem{rev2}
D. S. L. Abergel, V. Apalkov, J. Berashevich, K. Ziegler and T. Chakraborty, Adv. in Phys. \textbf{59}, 261 (2010).

\bibitem{rev3}
S. Das Sarma, Shaffique Adam, E. H. Hwang and Enrico Rossi, Rev. Mod. Phys. \textbf{83}, 109 (2011).

\bibitem{trans}
F. Schwierz, Nat. Nano. \textbf{5}, 487 (2010).

\bibitem{pereiraTBM}
V. M. Pereira, A. H. Castro Neto and N. M. R. Peres, Phys. Rev. B \textbf{80}, 045401 (2009).

\bibitem{confine}
M. Y. Han, B \"{O}zyilmaz, Y. Zhang, and P. Kim, Phys. Rev. Lett. \textbf{98}, 206805 (2007).

\bibitem{dope}
P. P. Shinde and V. Kumar, Phys. Rev. B \textbf{84}, 125401 (2011).

\bibitem{substrate}
D. Marchenko, A. Varykhalov, M. R. Scholz,	G. Bihlmayer, E. I. Rashba,	A. Rybkin, A. M. Shikin	and O. Rader, Nat. Comm. \textbf{3}, 1232, (2012).

\bibitem{BLG1}
E. McCann, D. S. L. Abergel and V. I. Fal’ko, Solid State Comm. \textbf{143}, 110 (2007).

\bibitem{BLG2}
E. McCann, D. S. L. Abergel and V. I. Fal’ko, Eur. Phys. J. Special Topics \textbf{148}, 91 (2007).

\bibitem{bias}
Y. Zhang, T.-T. Tang, C. Girit, Z. Hao, M. C. Martin, A. Zettl, M. F. Crommie, Y. R. Shen and F. Wang, Nature \textbf{459}, 820 (2009).

\bibitem{falko}
M. Mucha-Kruczy\'{n}ski, I. L. Aleiner and V. I. Fal’ko, Phys. Rev. B \textbf{84}, 041404(R) (2011).

\bibitem{conductance}
D. A. Gradinar, H. Schomerus and V. I. Fal’ko, Phys. Rev. B \textbf{85}, 165429 (2012).

\bibitem{peetersblg}
B. Verberck, B. Partoens, F. M. Peeters and B. Trauzettel, Phys. Rev. B \textbf{85}, 125403 (2012).

\bibitem{maria}
E. Mariani, A. J. Pearce and F. von Oppen, Phys. Rev. B \textbf{86}, 165448 (2012).

\bibitem{pereiraA}
A. L. Kitt, V. M. Pereira, A. K. Swan, and Bennett B. Goldberg, Phys. Rev. B \textbf{85}, 115432 (2012). 

\bibitem{error}
A. L. Kitt, V. M. Pereira, A. K. Swan, and Bennett B. Goldberg, Phys. Rev. B \textbf{87}, 159909(E) (2013).

\bibitem{peeters2}
M. Ramezani Masir, D. Moldovan and F. M. Peeters, Solid State Comm. \textbf{175}, 76 (2013).

\bibitem{domain}
J. A. Crosse, Phys. Rev. B \textbf{90}, 045201 (2014).

\bibitem{blgrev}
Edward McCann and Mikito Koshino, Rep. Prog. Phys. \textbf{76}, 056503 (2013).

\bibitem{saito}
R. Saito, G. Dresselhaus and M. S. Dresselhaus, \textit{Physical Properties of Carbon Nanotubes} (Imperial College Press, 1998).

\bibitem{poisson}
O. L. Blakslee, D. G. Proctor, E. J. Seldin, G. B. Spence, and T. Weng, J. Appl. Phys. \textbf{41}, 3373 (1970).

\bibitem{gamma}
Z. Q. Li, E. A. Henriksen, Z. Jiang, Z. Hao, M. C. Martin, P. Kim, H. L. Stormer and D. N. Basov, Phys. Rev. Lett. \textbf{102}, 037403 (2009).

\bibitem{sun}
D. Sun, Z.-K. Wu, C. Divin, X. Li, C. Berger, W. A. de Heer, P. N. First and T. B. Norris, Phys. Rev. Lett. \textbf{101}, 157402 (2008).

\bibitem{george}
P. A. George, J. Strait, J. Dawlaty, S. Shivaraman, M. Chandrashekhar, F. Rana and M. G. Spencer, Nano Lett. \textbf{8}, 4248 (2008).

\bibitem{physchem}
J. Zhang, K. P. Ong and P. Wu, J. Phys. Chem. C \textbf{114}, 12749 (2010).

\bibitem{low}
T. Low, Y. Jiang, M. Katsnelson and F. Guinea, Nano Lett. \textbf{12}, 850 (2012).

\bibitem{vanveen}
J. van Veen, A. Castellanos-Gomez, H. S. J. van der Zant, G. A. Steele, Graphene \textbf{2}, 13 (2013).

\bibitem{mohanty}
N. Mohanty, D. Moore, Z. Xu, T.S. Sreeprasad, A. Nagaraja, A. A. Rodriguez and V. Berry, Nat. Comm. \textbf{3}, 844 (2012).

\bibitem{youngs}
C. Lee, X. Wei, J. W. Kysar, J. Hone, Science \textbf{321}, 385 (2008).

\bibitem{pet}
T. M. G. Mohiuddin, A. Lombardo, R. R. Nair, A. Bonetti, G. Savini, R. Jalil, N. Bonini, D. M. Basko, C. Galiotis, N. Marzari, K. S. Novoselov, A. K. Geim and A. C. Ferrari, Phys. Rev. B \textbf{79}, 205433 (2009).

\bibitem{peeters}
D. Moldovan, M. Ramezani Masir and F. M. Peeters,  Phys. Rev. B \textbf{88}, 035446 (2013).

\bibitem{nanoindent}
M. Neek-Amal and F. M. Peeters, Phys. Rev. B \textbf{81}, 235421 (2010).
 
\bibitem{trilayer}
C. H. Lui, Z. Li, K. F. Mak, E. Cappelluti and Tony F. Heinz, Nat. Phys. \textbf{7}, 944 (2011).

\bibitem{fewlayer}
S. Bala kumar and J. Guo, Appl. Phys. Lett. \textbf{98}, 222101 (2011).

\bibitem{mayo}
A. S. Mayorov, D. C. Elias, M. Mucha-Kruczynski, R. V. Gorbachev, T. Tudorovskiy,
A. Zhukov, S. V. Morozov, M. I. Katsnelson, V. I. Fal’ko, A. K. Geim, K. S. Novoselov, Science \textbf{333}, 860 (2011).




\end{thebibliography}
\end{document}